\documentclass[trackchanges]{aastex63}

\received{***}
\revised{***}
\accepted{***}
\submitjournal{ApJS - Parker Solar Probe Special Issue}

\shorttitle{Temperature Anisotropy}
\shortauthors{Huang et al.}
\graphicspath{{./}{figures/}}

\begin{document}

\title{Proton Temperature Anisotropy Variations in Inner Heliosphere Estimated with First Parker Solar Probe Observations}

\correspondingauthor{Jia Huang}
\email{jiahu@umich.edu}

\author[0000-0002-9954-4707]{Jia Huang}
\affiliation{Climate and Space Sciences and Engineering, University of Michigan, Ann Arbor, MI 48109, USA}

\author[0000-0002-7077-930X]{J. C. Kasper}
\affiliation{Climate and Space Sciences and Engineering, University of Michigan, Ann Arbor, MI 48109, USA}
\affiliation{Smithsonian Astrophysical Observatory, Cambridge, MA 02138 USA.}

\author[0000-0003-1542-1302]{D. Vech}
\affiliation{Climate and Space Sciences and Engineering, University of Michigan, Ann Arbor, MI 48109, USA}
\affiliation{Laboratory for Atmospheric and Space Physics, University of Colorado, Boulder, CO, USA\\}

\author[0000-0001-6038-1923]{K. G. Klein}
\affiliation{Lunar and Planetary Laboratory, University of Arizona, Tucson, AZ 85719, USA.}

\author[0000-0002-7728-0085]{M. Stevens}
\affiliation{Smithsonian Astrophysical Observatory, Cambridge, MA 02138 USA.}

\author[0000-0002-7365-0472]{Mihailo M. Martinovi\'c}
\affiliation{Lunar and Planetary Laboratory, University of Arizona, Tucson, AZ 85719, USA.}
\affil{LESIA, Observatoire de Paris, Universite PSL, CNRS, Sorbonne Universite, Universite de Paris, 5 place Jules Janssen, 92195 Meudon, France}

\author[0000-0001-6673-3432]{B. L. Alterman}
\affiliation{Department of Applied Physics, University of Michigan, 450 Church St., Ann Arbor, MI 48109, USA}
\affiliation{Climate and Space Sciences and Engineering, University of Michigan, Ann Arbor, MI 48109, USA}

\author[0000-0003-4247-4864]{Tereza \v{D}urovcov\'a}
\affiliation{Faculty of Mathematics and Physics, Charles University, Prague, Czech Republic}

\author[0000-0002-5699-090X]{Kristoff Paulson}
\affiliation{Smithsonian Astrophysical Observatory, Cambridge, MA 02138 USA.}

\author[0000-0002-2229-5618]{Bennett A. Maruca}
\affiliation{Department of Physics and Astronomy, University of Delaware, Newark, DE 19716, USA}
\affil{Bartol Research Institute, University of Delaware, Newark, DE 19716}

\author[0000-0001-8358-0482]{Ramiz A. Qudsi}
\affiliation{Department of Physics and Astronomy, University of Delaware, Newark, DE 19716, USA}

\author[0000-0002-3520-4041]{A. W. Case}
\affiliation{Smithsonian Astrophysical Observatory, Cambridge, MA 02138 USA.}

\author[0000-0001-6095-2490]{K. E. Korreck}
\affiliation{Smithsonian Astrophysical Observatory, Cambridge, MA 02138 USA.}

\author[0000-0002-6849-5527]{Lan K. Jian}
\affiliation{Heliophysics Science Division, NASA Goddard Space Flight Center, Greenbelt, USA}

\author[0000-0002-2381-3106]{Marco Velli}
\affiliation{Department of Earth, Planetary and Space Sciences, University of California, Los Angeles CA 90095, USA}

\author[0000-0001-6807-8494]{B. Lavraud}
\affiliation{Institut de Recherche en Astrophysique et Planétologie, CNRS, UPS, CNES, Université de Toulouse, Toulouse, France}

\author[0000-0001-6247-6934]{A. Hegedus}
\affiliation{Climate and Space Sciences and Engineering, University of Michigan, Ann Arbor, MI 48109, USA}

\author[0000-0002-9694-174X]{C. M. Bert}
\affil{Climate and Space Sciences and Engineering, University of Michigan, Ann Arbor, MI 48109, USA}

\author{J. Holmes}
\affiliation{Climate and Space Sciences and Engineering, University of Michigan, Ann Arbor, MI 48109, USA}

\author[0000-0002-1989-3596]{Stuart D. Bale}
\affil{Physics Department, University of California, Berkeley, CA 94720-7300, USA}
\affil{Space Sciences Laboratory, University of California, Berkeley, CA 94720-7450, USA}
\affil{The Blackett Laboratory, Imperial College London, London, SW7 2AZ, UK}
\affil{School of Physics and Astronomy, Queen Mary University of London, London E1 4NS, UK}

\author[0000-0001-5030-6030]{Davin E. Larson}
\affiliation{University of California, Berkeley: Berkeley, CA, USA.}

\author[0000-0002-0396-0547]{Roberto Livi}
\affiliation{University of California, Berkeley: Berkeley, CA, USA.}

\author[0000-0002-7287-5098]{P. Whittlesey}
\affiliation{University of California, Berkeley: Berkeley, CA, USA.}

\author{Marc Pulupa}
\affiliation{University of California, Berkeley: Berkeley, CA, USA.}

\author{Robert J. MacDowall}
\affiliation{NASA Goddard SFC, Greenbelt, MD, USA}

\author[0000-0003-1191-1558]{David M. Malaspina}
\affiliation{Astrophysical and Planetary Sciences Department, University of Colorado, Boulder, CO, USA}
\affiliation{Laboratory for Atmospheric and Space Physics, University of Colorado, Boulder, CO, USA}

\author{John W. Bonnell}
\affiliation{University of Colorado, Boulder, Boulder, CO, USA}

\author{Peter Harvey}
\affiliation{University of Colorado, Boulder, Boulder, CO, USA}

\author{Keith Goetz}
\affiliation{University of Minnesota, Minneapolis, MN, USA}

\author[0000-0002-4401-0943]{Thierry Dudok de Wit}
\affiliation{LPC2E, CNRS and University of Orl\'eans, 3A avenue de la
Recherche Scientifique, Orl\'eans, France}



\begin{abstract}
We present a technique for deriving the temperature anisotropy of solar wind protons observed by the Parker Solar Probe mission in the near-Sun solar wind.   The variation in the temperature of solar wind protons in the radial direction measured by the SWEAP Solar Probe Cup is compared with variation in the orientation of the local magnetic field measured by the FIELDS fluxgate magnetometer, and the components of the proton temperature parallel and perpendicular to the magnetic field are extracted.  This procedure is applied to both moments of the proton velocity distribution function (VDF) and to the results of a non-linear fit of proton core and proton beam Maxwellian components of the VDF, and the results are compared and optimum timescales for data selection and trends in the uncertainty in the method are identified.  We find that the moment-based proton temperature anisotropy is more physically consistent with the expected limits of the mirror and firehose instabilities, possibly because the fits do not capture a significant non-Maxwellian shape to the proton VDF near the Sun.  Several radial trends in the temperature components and the variation of the anisotropy with parallel plasma beta are presented.  We find that the proton parallel plasma beta seen by PSP in the first encounter is not significantly smaller than the values seen by Helios further from the Sun, possibly as a result of long term variation in the heliospheric density and magnetic field over the last half-century.  The observed radial dependence of temperature variations in the fast solar wind implies stronger perpendicular heating and parallel cooling than previous results from Helios measurements made at larger radial distances, but a similar anti-correlation between proton temperature anisotropy and parallel plasma beta persists closer to the Sun. The temperature anisotropies of the slow solar wind are well constrained by the mirror and parallel firehose instabilities. The perpendicular heating of the slow solar wind inside 0.24 AU may contribute to it reaching the mirror instability threshold. These results suggest that we may see stronger anisotropic heating as PSP moves closer to the Sun, and that a careful treatment of the shape of the proton VDF may be needed to correctly describe the temperature.
\end{abstract}


\section{Introduction} \label{sec:intro}
Temperatures perpendicular ($T_{\perp p}$, see Appendix for definitions) and parallel ($T_{\parallel p}$) to the ambient magnetic field ($\mathbf{B}$) are one measure of the solar wind's departure from thermal equilibrium. Characterizing such departures is pivotal to understanding the kinetic processes governing the dynamics of interplanetary medium \citep{kasper-2002, kasper-2007, maruca-2012, he-2013, maruca-2013}. Temperature anisotropy ($T_{\perp p}/T_{\parallel p} \neq 1$) arises when anisotropic heating and cooling processes act preferentially in one direction \citep{maruca-2011}; such preferential heating is supported by observed departures of $T_{\perp p}/T_{\parallel p}$ from adiabatic predictions in solar wind observations \citep{matteini-2007}. For an ideal spherical adiabatic expansion with a polytropic index $\gamma = 5/3$, the total proton temperature is expected to decrease with heliocentric distance with an index of $T_p \propto  R^{-4/3}$. However, observations suggest a much slower decay rate, implying continual proton heating over extended radial distances \citep{hellinger-2011}. Moreover, assuming the collisionless solar wind has small heat fluxes and rare interactions \citep{kasper-2003, matteini-2013, perrone-2018}, the double-adiabatic equations of state predict that the adiabatic invariants $T_{\perp p}/B$ and $T_{\parallel p}B^2/n_{p}^2$ should be conserved \citep{chew-1956}. As $B$ and proton density $n_p$ decrease as $R^{-2}$ for a solar wind expanding with constant speed, $T_{\perp p}$ is expected to decrease with $R^{-2}$ and $T_{\parallel p}$ should be constant. However, Helios measurements covering radial distances from 0.3 AU to 1 AU \citep[e.g.][]{marsch-1982, marsch-1983, hellinger-2011} showed that these adiabatic invariants are not conserved, with $T_{\perp p}$ decreasing slower and $T_{\parallel p}$ decreasing faster than double-adiabatic predictions, implying both a preferential perpendicular heating and parallel cooling of protons \citep{hellinger-2011}.  Additionally $T_\perp/B$, effectively the magnetic moment of the solar wind protons, was found to increase with distance in fast wind instead of being conserved, further indicating that the plasma was being heated preferentially perpendicular to the magnetic field.

As temperature anisotropy departs from unity, anisotropy-driven instabilities such as mirror, ion-cyclotron, parallel and oblique firehose instabilities arise, and act to isotropize the plasma \citep{gary-1993, liu-2005, liu-2007, maruca-2011}. However, the thresholds of these instabilities are different in different conditions (e.g. solar wind, magnetosphere, magnetosheath), and many researchers have tested the constrains of these instabilities \citep[][and references therein]{kasper-2002}. \citet{gary-2000} studied the constraints from ion-cyclotron instability based on theoretical and numerical methods. \citet{kasper-2003} found that the dominant limit in the large plasma $\beta$ regime for $T_{\perp p}>T_{\parallel p}$ transitions from the ion-cyclotron to the mirror instability threshold. Using Wind measurements of the solar wind, \citet{kasper-2002} demonstrated that the firehose instability serves as a constraint on the proton temperature anisotropies when $T_{\perp p}<T_{\parallel p}$. An extended work using Wind data by \citet{hellinger-2006} argued that the oblique instabilities (mirror and oblique firehose instabilities) more effectively constrain the proton temperature anisotropy for slow solar wind, while the mirror and parallel firehose instabilities probably play a role on limiting the proton core temperature anisotropy for fast wind. Besides, both Wind data from 1 AU and Helios data close to 0.3 AU exhibit an anti-correlation between $\beta_{\parallel p}$ and $T_{\perp p}/T_{\parallel p}$ for the proton core population \citep{marsch-2004,hellinger-2006}. Using Nyquist's instability criterion and by assessing ion sources of free energy, \citet{klein-2017,klein-2018} found that instabilities are pervasive in the solar wind rather than simply serving as a boundary.

The above results are drawn exclusively from measurements at distances greater than 0.3 AU. It is valuable to include near-Sun observations to comprehensively and thoroughly investigate how the temperature components and adiabatic invariants varying with distances from the Sun, and to study the impact of anisotropy-drive instabilities in the solar wind at radial distances below 0.3 AU. Parker Solar Probe (PSP) \citep{fox-2016} is designed to fly into the solar atmosphere, reaching a deepest perihelion at $\sim 9.8$ solar radii ($R_S$) at the end of the mission. Currently, PSP has operated through several encounters with the Sun at its initial perihelion of $35.7 R_S$, and initial overviews of the solar wind plasma seen during these encounters have been reported \citep{Kasper-2019, Bale-2019}.  As such it provides novel data that can constrain the behavior and significance of instabilities to solar wind evolution. In this work, we derive temperature anisotropies from PSP observations during the first solar encounter (E1). We present the data and methodology in Section \ref{sec:data} and Section \ref{sec:method}, respectively. Section \ref{sec:tdis} shows the temperature variations with distance from the Sun, which may indicate stronger perpendicular heating and parallel cooling effects than previous results. Section \ref{sec:tbeta} presents the temperature anisotropy variations with $\beta_{\parallel p}$, suggesting that the mirror and parallel firehose instabilities may well constrain the temperature anisotropy of slow solar wind. We summarize our results in Section \ref{sec:summary}, and present the details of our fitting technique used to derive temperature anisotropies in Appendices A-D.

\section{Data} \label{sec:data}
PSP carries the Solar Wind Electrons, Alphas, and Protons (SWEAP) instrument suite \citep{kasper-2016} and the FIELDS instrument suite \citep{bale-2016}. SWEAP is a thermal ion package designed to measure velocity distributions of solar wind electrons, alpha particles, and protons. The suite includes the Solar Probe Cup (SPC) \citep{Case-2019} and Solar Probe Analyzers (SPANs) \citep{Whittlesey-2019, Livi-2019}. SPC is a Sun-pointed Faraday cup (FC). The SPANs have an A and B component, each consisting of one or more electrostatic analyzers (ESAs). SPAN-A is mounted on the ram side and includes an ion and electron ESA. SPAN-B contains an electron ESA mounted on the anti-ram side. In this paper, we focus on proton measurements derived from SPC. SPC reports proton measurements derived from both \textit{moment} and \textit{non-linear} fitting algorithms. The \textit{moment} algorithm returns a single, isotropic proton population. The \textit{non-linear} fitting algorithm returns a proton core and a proton beam population. Typically, the proton core corresponds to the peak of the solar wind proton velocity distribution function (VDF) and the beam corresponds to its shoulder. A summed core+beam population by taking into account of their relative drift is also reported. Future SWEAP data products will include measurements of the proton temperature anisotropy using the three dimensional VDF seen by SPAN-A, but this analysis is under development and PSP has yet to achieve a sufficiently large orbital velocity for the peak of the solar wind VDF to be seen by SPAN-A. FIELDS is designed to measure DC and fluctuation magnetic and electric fields, plasma wave spectra and polarization properties, the spacecraft floating potential, and solar radio emissions \citep{bale-2016}.

SPC's operation mode varies with distance from the Sun. During near-Sun encounters ($R < 0.25 \, \mathrm{AU}$ or $54 \, R_S$), its sampling rate is highest \citep{kasper-2016}. The Encounter mode collected one measurement every $0.874$ seconds in E1 \citep{Case-2019}. For non-encounter cruise operations, time resolution is lowered. The Cruise mode collected one solar wind VDF every $27.962$ seconds \citep{Case-2019} during E1. We select intervals for which all the SPC proton quality flags (excepting the four flags associated with helium measurements that are still under calibration) indicate good observations, and the meaning of each quality flag is explained in \citet{Case-2019}.

PSP/FIELDS collects high resolution vector magnetic fields with variable time resolution. During E1, the data rates range between $2.3 \, \mathrm{Hz}$ to $293 \, \mathrm{Hz}$ \citep{Bale-2019}. As these data rates are markedly higher than SPC's, we down-sample the time resolution to that of plasma data for this work.

\section{Method} \label{sec:method}
As noted above, there are two techniques to derive plasma parameters from FC measurements, one is \textit{non-linear} fitting technique and the other is a summed \textit{moment} technique \citep{kasper-2002thesis, kasper-2006}. \citet{kasper-2002thesis} suggests that the \textit{non-linear} fitting technique provides far more information than the \textit{moment} algorithm, but the \textit{moment} algorithm is used due to its simplicity, and it provides an easy visualization of the temperature anisotropies.

An anisotropic, magnetized plasma has different thermal speeds parallel ($w_\parallel$) and perpendicular ($w_\perp$) to the local magnetic field. Because a FC measures the reduced VDF and has a very uniform angular response, the instrument reports an effective thermal speed ($\tilde{w}$) that is a function of the orientation between the FC's look direction and the ambient magnetic field \citep{kasper-2002thesis, kasper-2002, kasper-2006}. For FC look direction $\hat{n}$ and magnetic field direction $\hat{b}$, $\tilde{w}$ is given by:

\begin{equation}
\label{eq:wtilde}
\tilde{w} = \sqrt{w_\parallel^2 \left(\hat{n}\cdot \hat{b} \right)^2 + w_\perp^2 \left(1 - \left(\hat{n} \cdot \hat{b}\right)^2\right)}
\end{equation}.

\citet{kasper-2002thesis} applies this equation to Wind FC data, which utilize multiple look directions within a single measurement. In contrast, SPC only utilizes a single look direction. During encounter phases of the orbit, this direction is Sun-pointed, i.\ e.\ $\hat{n} = \hat{r}$. As such, we can replace $\left(\hat{n} \cdot \hat{b}\right)^2$ with $\left(\hat{r} \cdot \hat{b}\right)^2 = \left(B_r/B\right)^2$. To apply the Wind/FC techniques, we treat successive radial measurements as if they were the multiple FC look directions and combine them with magnetic fields measurements to extract a temperature anisotropy. The Appendix covers the details of our algorithm. Hereafter, our results and analysis utilize the selected temperature anisotropy measurements for total proton population that derived from E1 \textit{moment} data covering the dates October 20\textsuperscript{th}, 2018 to November 24\textsuperscript{th}, 2018. We bin the data with a 1-minute rolling boxcar that steps with 10-second increments and then fit the data in each bin. For the Cruise data, we implement a 4-minute boxcar with a 1-minute moving step. In both cases, the moving step corresponds to the frequency of resulting anisotropy measurements. During this time period, SPC has 1,787,558 measurements in total, of which 1,629,944 (91.2\%) are flagged as good observations. The majority of the measurements (1,459,029, 89.5\%) are in slow solar wind (SSW, $v_{sw}\leq450 \ km \ s^{-1}$). The remaining 170,915 (10.5\%) measurements are in fast solar wind (FSW, $v_{sw}>450 \ km \ s^{-1}$). Our boxcar algorithm transforms these spectra into 129,586 anisotropy measurements in total, of which 78,499 (60.6\%) meet the data quality selection criteria outlined in Appendix \ref{sec:select}. Among them, slow solar wind dominates 86.2\% (67,668 fittings), and fast solar wind dominates the rest 13.8\% (10,831 fittings).

\begin{figure}
\epsscale{0.8}
\plotone{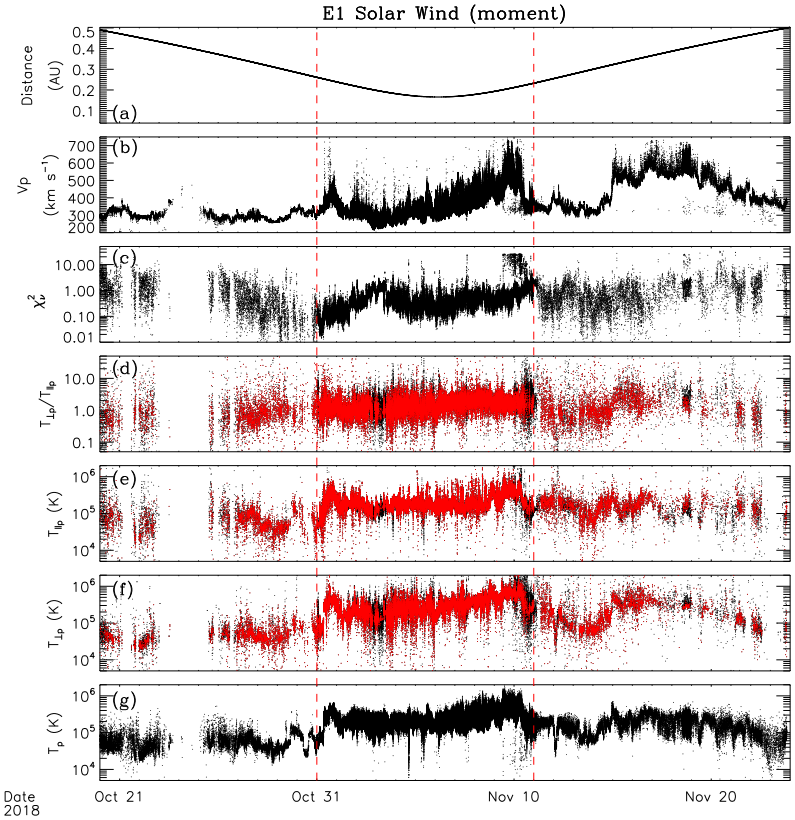}
\caption{Overview of temperature variations during Encounter 1. From top to bottom, the panels show the spacecraft distance from the Sun, solar wind speed, the goodness of fit $\chi_{\nu}^2$, temperature anisotropy, parallel temperature, perpendicular temperature, and total temperature. Red points in panel (d) to (f) indicates good data points selected by our selection criteria as shown in Appendix \ref{sec:select}. The dashed vertical lines show high time resolution data inside 0.25 AU. }. \label{fig:overview}
\end{figure}

Figure \ref{fig:overview} shows an overview of proton temperature variations during E1, including both Encounter and Cruise mode data. High time resolution data inside 0.25 AU are from October 31\textsuperscript{th} to November 11\textsuperscript{th} as shown by the dashed vertical lines, and low time resolution data cover heliocentric distance from 0.25 AU to about 0.5 AU. \citet{Kasper-2019} found that protons are 3 to 4 times hotter than protons with similar solar wind speed at 1 AU. Panel (a) and (g) are consistent with their results, showing a decrease in $T_p$ with increasing distance from the Sun. Panel (b) shows the bulk solar wind speed. $\chi_{\nu}^2$ (reduced-$\chi^2$ or $\chi^2$ per degree of freedom) in Panel (c) measures the goodness of fit, and the fitting is good when this parameter approaches unity \citep{bevington-1993}. Panel (d) presents the temperature anisotropy. The parallel and perpendicular temperatures in Panel (e) and Panel (f) show similar variations as the total temperature. The red (black) points in Panels (d) through (f) indicate data points that are (not) selected with the criteria present in Appendix \ref{sec:select}.

\section{Temperature variations with distance from the Sun \label{sec:tdis}}
\begin{figure}
\epsscale{1.0}
\plotone{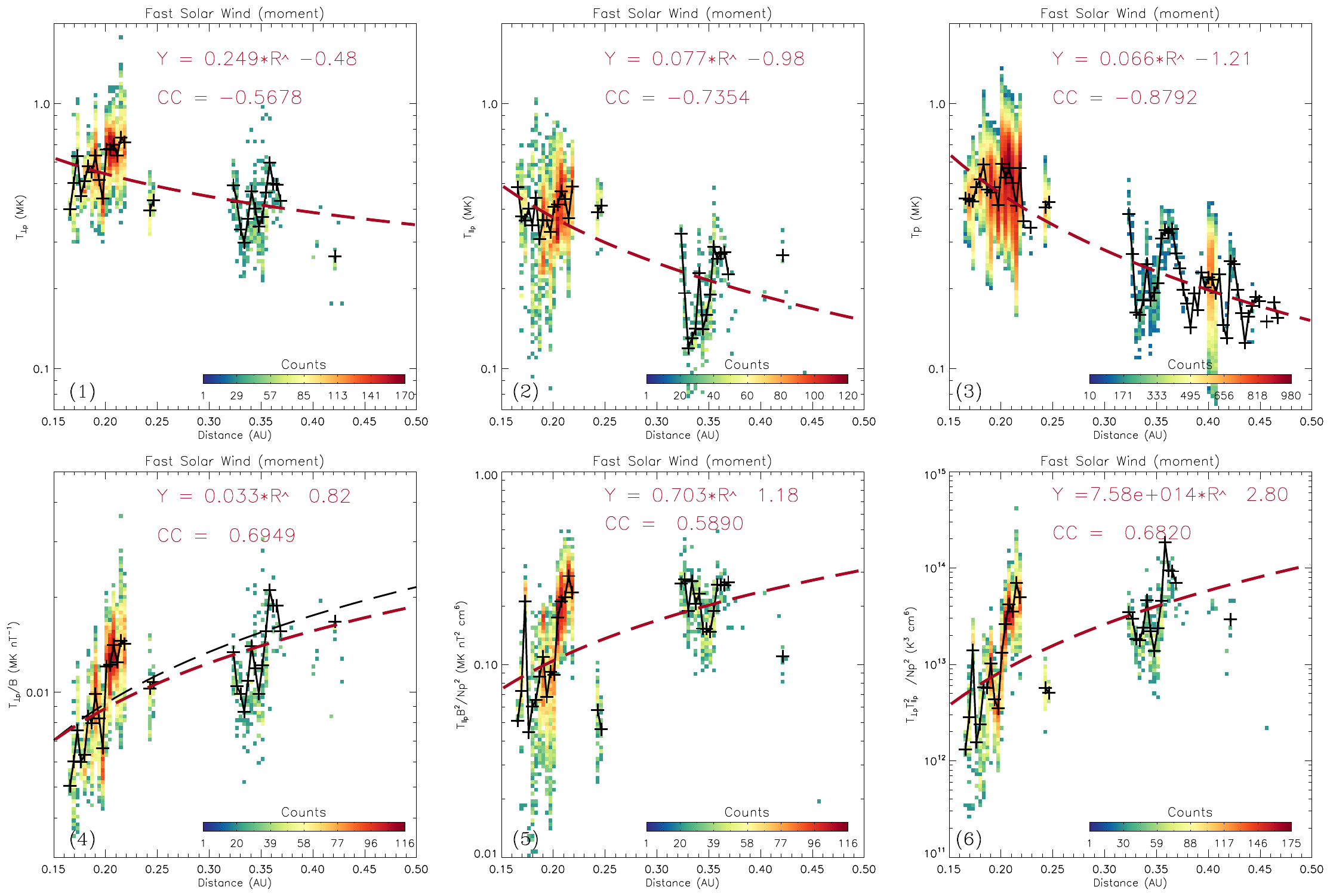}
\caption{Temperature variations with heliocentric distance for fast solar wind. The upper panels show perpendicular temperature $T_{\perp p}$, parallel temperature $T_{\parallel p}$ and total temperature $T_{p}$, and the lower panels show the double-adiabatic invariants $T_{\perp p}/B$, $T_{\parallel p}B^2/N_{p}^2$, and invariant $T_{\parallel p}T_{\perp p}^2/N_{p}^2$. Color indicates observation counts. The black crosses in each panel are average values at different bins of heliocentric distance, and the value is calculated when the bin includes at least 30 measurements. Red dashed lines represent the fitted relationships, with the fitted parameters and correlation coefficients present. In Panel (4), the black dashed line indicates the Helios results adapted from \citet{tu-1988} and \citet{marsch-1991}. }. \label{fig:tdistance}
\end{figure}

In this section, we investigate the radial variations of temperature components ($T_{\perp p}$, $T_{\parallel p}$ and $T_{p}$) and adiabatic invariants ($T_{\perp p}/B$, $T_{\parallel p}B^2/n_{p}^2$, and $T_{\parallel p}T_{\perp p}^2/n_{p}^2$) in different solar wind conditions. Take fast solar wind as an example, Figure \ref{fig:tdistance} presents the results for solar wind with bulk speed larger than 450 $km \ s^{-1}$, with the color and black crosses in each panel indicating the measurement counts and average values for each bin, and the red dashed line representing the fitted radial evolutions. Here, we use a linear fitting method to fit the logarithm values of both the parameter and heliocentric distance (the later in AU), and present the relationship in exponential format $Y = a (R/R_0)^b$, where $R_0$ equals 1 AU, $a$ is the 1 AU intercept, and $b$ is the power-law index. In Panel (4), we overlap $T_{\perp p}/B$ variations for solar wind speed ranges from 600 to 700 $km \ s^{-1}$ with Helios data between 0.3 AU and 1 AU (black dashed line, the arbitrary coefficient is adapted for convenience to make comparison), indicating good agreement between both results.

Table \ref{tab:Tvsdis} lists the fitted radial evolution indexes with one-sigma uncertainty for each parameter in fast solar wind (FSW), slow solar wind (SSW), inbound orbit solar wind and outbound orbit solar wind, respectively. In order to reduce the effect of measurements numbers at different distances, we fit the data with mean values for each bin (e.g. the black crosses in Figure \ref{fig:tdistance}), and we calculate the mean value only when the bin includes at least 30 measurements. For comparison, we list the indexes from \citet{marsch-1983}, \citet{hellinger-2011} and \citet{perrone-2018} , which are derived from Helios observations between 0.3 AU and 1 AU and these works generally focus on fast solar wind with speed larger than 600 $km\ s^{-1}$. The inbound orbit predominantly observes slow solar wind, while the outbound orbit mainly observes fast solar wind. Therefore, the indexes for inbound (outbound) orbit and slow (fast) solar wind are somewhat similar.

Focusing on fast solar wind, we compare the indexes derived from PSP data and from Helios data. The total magnetic field strength $B$ decreases with an index of about -1.66, which is nearly the same as Helios observations. However, the plasma density $n_p$ decreases with an index of -2.59, which is much steeper than previous results and the adiabatic expansion index, implying non-spherical expansion geometry or more dynamic interactions closer to the Sun. For total temperature $T_{p}$, the index is -1.21, slightly steeper than Helios results. The three parameters have small errors.
From this table, it seems the perpendicular temperature $T_{\perp p}$ and parallel temperature $T_{\parallel p}$ show significant deviations from double-adiabatic theory predictions when close to the Sun. $T_{\perp p}$ decreases with an index of -0.48, about two times slower than Helios observations, while $T_{\parallel p}$ decreases with an index of -0.98, which is about two times faster than previous results. As stated above, theory predicts $T_{\perp p}$ to decrease with $R^{-2}$ and $T_{\parallel p}$ to be conserved, the much larger deviations may imply more significant perpendicular heating and parallel cooling processes inside 0.3 AU. \citet{marsch-1983} found the so-called double-adiabatic invariants are broken because of possible wave-particle interaction or Coulomb collisions. It is not surprising that PSP also observes the same signature. In comparison, $T_{\perp p}/B$ evolves slightly faster with an index of 0.82, $T_{\parallel p}B^2/N_{p}^2$ increases much faster than Helios measurements with the index 1.18 versus 0.30, and the consequent invariants $T_{\parallel p}T_{\perp p}^2/N_{p}^2$ reveals the combined difference of the two double-adiabatic invariants. The even more prominent signatures could be caused by the faster decreases of plasma density, and the more remarkable deviations of $T_{\perp p}$ and $T_{\parallel p}$ evolution processes. We note the estimated errors for these parameters are large because the fact that we have less measurements beyond 0.25 AU, and our selection criteria also exclude some data points there. With more data in the future, we will be able to better constrain these radial trends.

In contrast to some earlier studies, we use 450 $km \ s^{-1}$ instead of 600 $km \ s^{-1}$ to select fast solar wind because PSP observes few solar wind above 600 $km \ s^{-1}$ (about 0.6\% of good measurements) except in spikes during E1 \citep{Kasper-2019}, which may also contribute to the discrepancies in Table \ref{tab:Tvsdis}. However, the fast solar wind we selected is mainly from E1 outbound orbit, and the solar wind speed increases with distances as shown in Panel (b) of Figure \ref{fig:overview}. Thus, we are probably safe to use 450 $km \ s^{-1}$ when close to the Sun to select fast wind in this work. We also note that the outbound orbit solar wind shows much similar power-law indexes as Helios fast solar wind. However, PSP observes fast, slow and fast solar wind as it leaves perihelion. Thus, the indexes for outbound solar wind are results of the mix of fast and slow solar wind, implying it will be valuable to investigate the evolution of indexes with different solar wind speed criterion in the future.

\begin{deluxetable}{ccccccc}
\tablecaption{Radial variation indexes for temperature components and adiabatic invariants. \label{tab:Tvsdis}}
\tablecolumns{10}
\tablenum{1}
\tablewidth{5pt}
\tablehead{
\colhead{Parameter} & \multicolumn{4}{c}{PSP Data}& \multicolumn{2}{c}{Helios Data} \\
\colhead{} & \colhead{Inbound} & \colhead{Outbound} & \colhead{SSW\tablenotemark{a}} &
\colhead{FSW\tablenotemark{b}} &\colhead{Perrone+2018\tablenotemark{c}} &
\colhead{Marsch+1983 \& Hellinger+2011 \tablenotemark{d}}
}
\startdata
$T_{p}$                                 &  -1.45$\pm$0.16  &  -0.90$\pm$0.15  &  -1.34$\pm$0.12 & -1.21$\pm$0.09  &  -0.90$\pm$0.08  &  -0.74                                     \\
$T_{\perp p}$                           &  -2.07$\pm$0.17  &  -0.94$\pm$0.27  &  -1.94$\pm$0.15 & -0.48$\pm$0.13  &  -0.99$\pm$0.08  &  -0.83                                     \\
$T_{\parallel p}$                       &  -1.19$\pm$0.21  &  -0.43$\pm$0.16  &  -0.99$\pm$0.13 & -0.98$\pm$0.16  &  -0.48$\pm$0.09  &  -0.54                                     \\
$T_{\perp p}/B$                         &   0.14$\pm$0.17  &	 0.68$\pm$0.24  &   0.07$\pm$0.15 &  0.82$\pm$0.15  &   0.65$\pm$0.08  &   0.60$\pm$0.90 \ (300-800 \ $km \ s^{-1}$)  \\
$T_{\parallel p}B^2/N_{p}^2$            &  -2.08$\pm$0.44  &   0.52$\pm$0.67  &  -1.79$\pm$0.37 &  1.18$\pm$0.29  &   0.30$\pm$0.20  &  -0.35$\pm$0.18 \ (400-500 \ $km \ s^{-1}$)  \\
                                        &                &                &               &               &                &  -0.58$\pm$0.19 \ (500-600 \ $km \ s^{-1}$)  \\
$T_{\parallel p}T_{\perp p}^2/N_{p}^2$  &  -2.69$\pm$0.65  &	 1.56$\pm$1.08  &  -1.88$\pm$0.65 &  2.80$\pm$0.54  &   1.60$\pm$0.30  &   0.60$\pm$1.20 \ (400-600 \ $km \ s^{-1}$)  \\
                                        &                &                &               &               &                &   1.8         \ (300-400 \ $km \ s^{-1}$)  \\
$N_{p}$                                 &  -1.94$\pm$0.11  &  -2.44$\pm$0.17  &  -2.38$\pm$0.28 & -2.59$\pm$0.18  &  -2.02$\pm$0.05  &  -1.8                                      \\
$B$                                     &  -1.90$\pm$0.07  &  -1.59$\pm$0.06  &  -1.83$\pm$0.06 & -1.66$\pm$0.06  &  -1.63$\pm$0.03  &  -1.6                                      \\
\enddata
\tablenotetext{a}{SSW: slow solar wind with speed $< 450 \ km \ s^{-1}$}
\tablenotetext{b}{FSW: fast solar wind with speed $> 450 \ km \ s^{-1}$}
\tablenotetext{c}{Perrone+2018: \citet{perrone-2018}, they focus on fast solar wind ($> 600 \ km\ s^{-1}$) observed by Helios spacecraft.}
\tablenotetext{d}{Hellinger+2011: \citet{hellinger-2011} focus on fast solar wind ($> 600 \ km \ s^{-1}$) observed by Helios spacecraft, including indexes for $T_{p}$, $T_{\perp p}$, $T_{\parallel p}$, $N_{p}$ and $B$; Marsch+1983: \citet{marsch-1983} provide indexes for  $T_{\perp p}/B$, $T_{\parallel p}B^2/N_{p}^2$ and $T_{\parallel p}T_{\perp p}^2/N_{p}^2$. }
\end{deluxetable}

\section{Temperature anisotropy variations with plasma beta \label{sec:tbeta}}
\begin{figure}
\epsscale{1.2}
\plotone{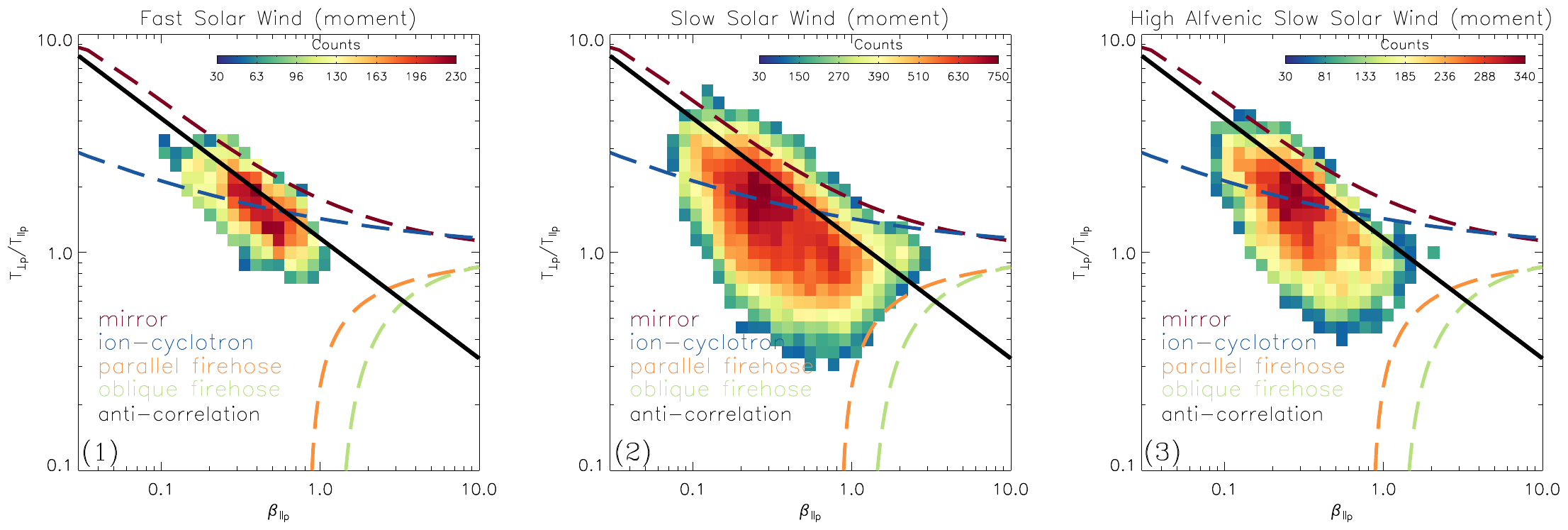}
\caption{Temperature anisotropy versus parallel plasma beta in different solar wind. Panel (1) to (3) shows their distributions for fast solar wind, slow solar wind and high Alfv\'enic slow solar wind, respectively. The dashed lines are colored to indicate different instabilities with thresholds from \citet{hellinger-2006}, and the solid line indicates the anti-correlations derived by \citet{marsch-2004}.}. \label{fig:tanibeta}
\end{figure}

As anisotropy-driven instabilities limit the departures of temperature anisotropy from unity, it is valuable to investigate whether they still work in inner heliosphere. In Figure \ref{fig:tanibeta}, we present the temperature anisotropy ($T_{\perp p}/T_{\parallel p}$) versus parallel plasma beta ($\beta_{\parallel p}$) for different types of solar wind. The red, blue, orange and green dashed lines represent mirror, ion-cyclotron, parallel firehose and oblique firehose instabilities, respectively, using the anisotropy-beta inverse relation from \citet{hellinger-2006}. The black solid line in each panel indicates the anti-correlation relationship between $T_{\perp p}/T_{\parallel p}$ and $\beta_{\parallel p}$ of proton core population, which was first derived from fast solar wind with Helios observations by \citet{marsch-2004}.

Due to the slow solar wind dominance during E1, the fast solar wind observations in Panel (1) only retain a small number of observations (13.8\%), however, the anti-correlation matches pretty well with both Wind and Helios results. Slow solar wind in Panel (2) reveals a large spread of data points, but their temperature anisotropies seem to be well constrained by the mirror and parallel firehose instabilities. It is different from the case of slow solar wind at 1 AU where mirror and oblique firehose instabilities work more effectively \citep{hellinger-2006}. This could be caused by the use of data from only one PSP encounter, and further investigation of the competition between parallel and oblique firehose instabilities when $T_{\perp p} < T_{\parallel p}$ is also needed. Moreover, Alfv\'enic slow solar wind is prevalent in the inner heliosphere, thus we also present the variations of high Alfv\'enic (normalized cross helicity $|\sigma_C|>0.7$) slow solar wind in Panel (3). It seems high Alfv\'enic slow wind has more anisotropic population than regular slow wind, which is similar to fast solar wind. However, the distribution shape may not deviate significantly from slow wind as shown in Panel (2).

\begin{figure}
\epsscale{1.2}
\plotone{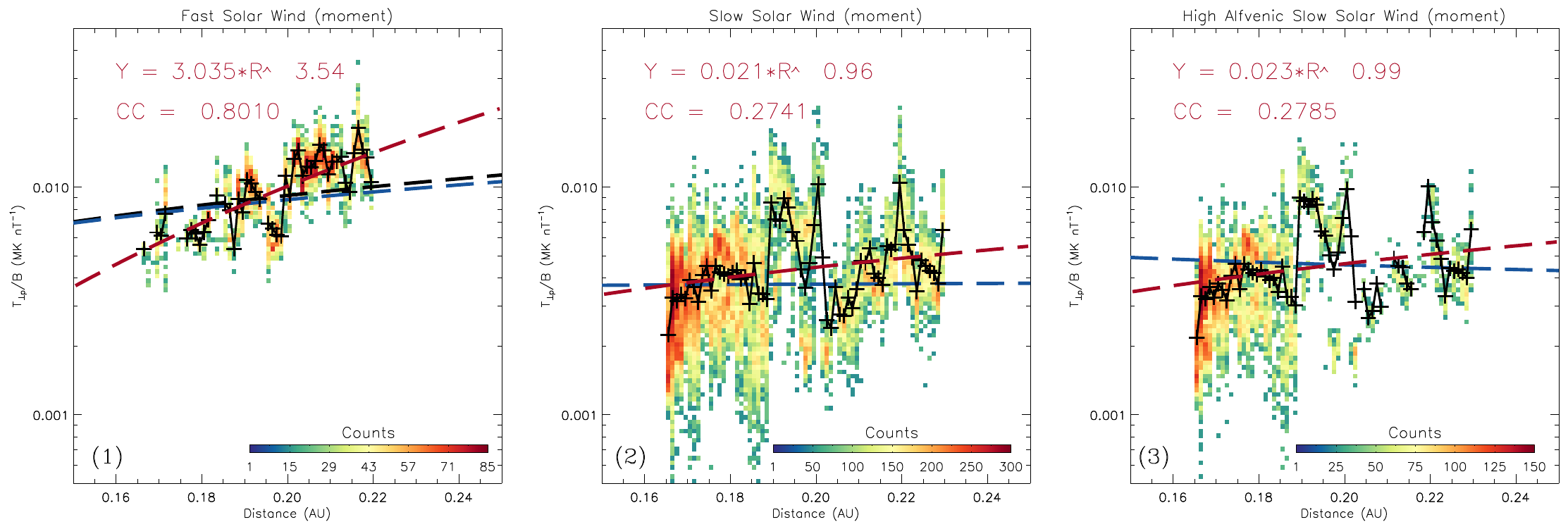}
\caption{Radial variations of magnetic moment $T_{\perp p}/B$ inside 0.24 AU for different solar wind. Panel (1) to (3) shows the variations for fast solar wind, slow solar wind and high Alfv\'enic slow solar wind, respectively. The format is similar to Figure \ref{fig:tdistance}. The red and blue dashed lines represent fitting results inside 0.24 AU and inside 0.50 AU, respectively. The black dashed line in Panel (1) indicates Helios results that adapted from \citet{tu-1988} and \citet{marsch-1991}. }. \label{fig:Tzoomin}
\end{figure}

Figure \ref{fig:tanibeta} further suggests that both slow and fast solar winds show similar trends up against the mirror instability thresholds, while in Helios only fast wind does \citep[e.g.][]{matteini-2007, matteini-2013}. Generally, more Coulomb collisions could wash out the anisotropy of slow solar wind plasma \citep[e.g.][]{marsch-1983b, kasper-2008}, thus the result may imply that PSP flies closer into the region where both the slow wind and fast wind experience anisotropic heating. Figure \ref{fig:Tzoomin} zooms in on the radial variations of magnetic moment $T_{\perp p}/B$ inside 0.24 AU. Similar to Figure \ref{fig:tdistance}, we present the fitting results inside 0.24 AU with red lines, and overlap the blue curves that derived from inside 0.5 AU. Panel (1) shows that the radial evolution in a longer distances (blue line) matches well with Helios results (black line) for fast solar wind, but fast wind inside 0.24 AU experiences much stronger perpendicular heating, with the magnetic moment increasing to an index of 3.5. For slow solar wind as shown in Panel (2) and (3), we also see stronger perpendicular heating compared to outside 0.24 AU, but weaker than for the fast solar wind, which is consistent with our expectations. Instead of the slow wind magnetic moment being essentially flat with distance, from $35-52 R_S$ it is observed to increase nearly linearly with distance. This is much weaker than the rapid increase in the fast solar wind magnetic moment, but it suggests that slow wind is also experiencing perpendicular heating close to the Sun.  Both regular slow solar wind and Alfv\'enic slow solar wind show a magnetic moment growing with distance, suggesting that the presence of intense Alfv\'en waves in the fast wind and the highly Alfv\'enic slow wind are not solely required for perpendicular heating.

\section{Conclusion \label{sec:summary}}
In this work, by applying linear fitting technique to the variation of the radial temperature of solar wind as a function of the direction of the magnetic field, we present the first estimation of temperature anisotropy with Parker Solar Probe E1 observations in the inner heliosphere. We find the radial perpendicular and parallel temperature in fast solar wind decreases with power-law indexes of -0.48 and -0.98, respectively. Comparing with Helios results, we suggest that PSP may observe more significant perpendicular heating and parallel cooling processes at distances between 0.5 and 0.166 AU. The prominent perpendicular heating could be mainly contributed by the stochastic heating process as confirmed by \citet{Martinovic-2019}. The fast solar wind also reveals strong anti-correlation between temperature anisotropy and parallel plasma beta. However, the mirror and parallel firehose instabilities seem to work as upper and lower limits to  constrain the temperature anisotropy of slow solar wind with E1 measurements. The perpendicular heating of slow solar wind inside 0.24 AU may contribute to the same trend against mirror instability thresholds as for fast solar wind. The high Alfv\'enic slow solar wind may not deviate significantly from regular slow solar wind.

\appendix
\section{Parameters} \label{sec:parameters}
Subscripts  ${\perp}$ and  ${\parallel}$ represent the perpendicular and parallel directions with respect to ambient magnetic field \textbf{B}. $T_{p}$,  $n_p$, $B_r$ and $R$ are the total proton temperature, proton number density, the radial component of magnetic field, and heliocentric distance. $T_{\perp p}$ and $T_{\parallel p}$ are the perpendicular and parallel proton temperatures, respectively. $T_{\perp p}/T_{\parallel p}$ is the proton temperature anisotropy, and $\beta_{\parallel p} = 2\mu_0 n_p k_B T_{\parallel p}/B^2$ is the parallel plasma beta, where $\mu_0$ and $k_B$ denotes the vacuum magnetic permeability and Boltzmann constant. We use $\chi_{\nu}^2$, i.e. reduced-$\chi^2$, to measure fitting goodness, which implies good fitting when it approaches to 1 \citep{bevington-1993}.

\section{Selection method} \label{sec:select}
Due to the SPC data having a variable time resolution, we need to find the best fitting window and moving step (i.e. the time resolution of fitted temperature anisotropy data) for high and low time resolution data, respectively.

\begin{figure}
\epsscale{1.0}
\plotone{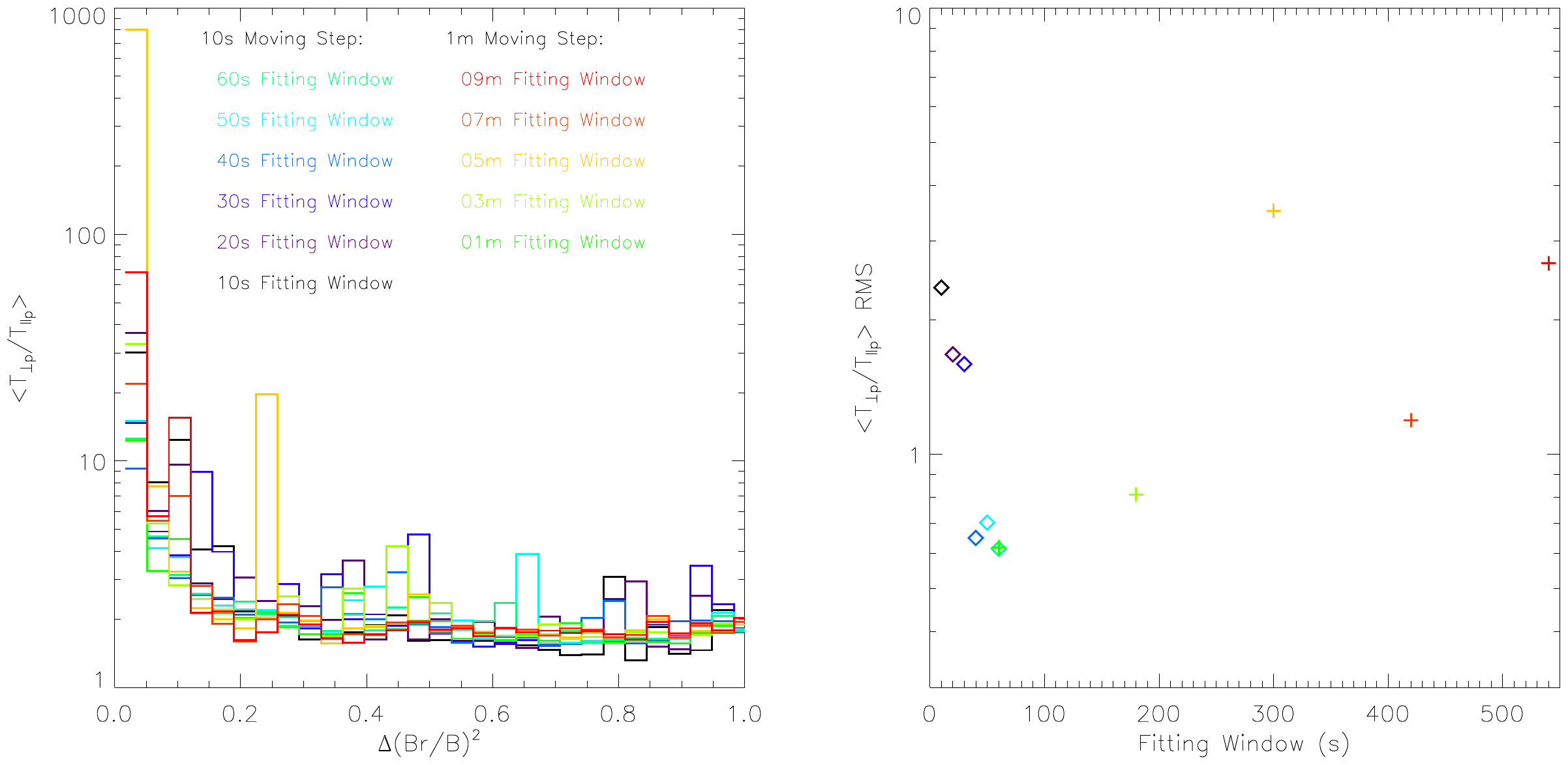}
\caption{Fitting goodness comparison. Left figure shows the average temperature anisotropy ( $\langle T_{\perp p}/T_{\parallel p}  \rangle$) variations with $\Delta(B_r/B)^2$, and different color represents data set derived from different fitting window and moving step. Right figure shows the root mean square (RMS) of average temperature anisotropy for each data set as indicated by the color, with the diamonds indicating 10-second moving step data sets and plus signs indicating 1-minute moving step data sets. }. \label{fig:fitcomp}
\end{figure}

We first choose high time resolution data from Oct. 31\textsuperscript{th}, 2018 to Nov. 11\textsuperscript{th}, 2018 for study. We derive eleven data sets for comparison. Six data sets are of 10-second time resolution, and the fitting window changes from 10-second, 20-second, 30-second, ..., to 60-second. The other five data sets are derived with 1-minute moving step, and the fitting window includes 1-minute, 3-minute, 5-minute, 7-minute and 9-minute. For each data set, if we have sufficient fittings and if we have selected the best time scale to fit the data, then the fitting results should be independent from the spread of data points ($\Delta(B_r/B)^2$) in each fitting window according to Equation \ref{eq:wtilde}. Thus, a uniform distribution of mean temperature anisotropy values ($\langle T_{\perp p}/T_{\parallel p}  \rangle$) with $\Delta(B_r/B)^2$ is expected. We have 95,034 fittings for data sets with 10-second moving step and 34,552 fittings for those with 1-minute moving step. The left histogram plot in Figure \ref{fig:fitcomp} shows the $\langle T_{\perp p}/T_{\parallel p}  \rangle$ variations with $\Delta(B_r/B)^2$ in 30 bins for the eleven data sets, and it is obvious that some data sets show more uniform distributions. However, $\langle T_{\perp p}/T_{\parallel p} \rangle$ value is much larger for each data set when $\Delta(B_r/B)^2$ is very small, which is reasonable because the fitting result could be arbitrary when the data points significantly concentrate together, implying a minimum $\Delta(B_r/B)^2$ should be applied to select good fittings. In order to find the most uniform distributions, we compare the root mean square value of $\langle T_{\perp p}/T_{\parallel p}  \rangle$ in 30 bins for each data set in right figure, and we exclude the unusual large value in the first histogram bin during calculations. The results predominantly show a decrease trend for 10-second time resolution data sets (diamonds) and an increase trend for 1-minute time resolution data sets (plus signs), and the minimum values for the two different time resolution data sets meet at 1 minute fitting window. Therefore, 1 minute fitting window could be the best. This suggests that 1 minute may be a good time scale to estimate temperature anisotropy before solar wind condition changes. Then, we use 1-minute fitting window to derive six data sets with moving step changes from 10-second to 60-second. According to our analysis (not shown), there is no significant difference between the fitting results, which may also imply that we have probably selected a good time scale to estimate the temperature anisotropy. We would like to set moving step as 10-second to increase the time resolution of temperature anisotropy data, and it is possible to further increase its time resolution. Consequently, we select the data set fitted with 1-minute window and 10-second moving step for further study. For the specific data set, we further require $\chi_{\nu}^2$ smaller than 1 and $\Delta(B_r/B)^2$ larger than, at least, 0.05 to exclude bad fittings. We can use more strict thresholds to select the data, but it seems the results are not significantly affected.

The Cruise science mode data have a lower time resolution of 27.962 seconds, i.e. two measurements in one minute, so we need to use a larger fitting window to include enough data points, but also choose a similar time scale as that we use to fit Encounter science mode data. We require the fitting window to include at least six data points, and at least two data points are changed when it moves to next step. Thus, for the low time resolution data from Oct. 20\textsuperscript{th}, 2018 to Oct. 31\textsuperscript{th}, 2018 and from Nov. 11\textsuperscript{th}, 2018 to Nov. 24\textsuperscript{th}, 2018, we use 1-minute moving step to derive nine data sets with the fitting window varies from 4-minute, 6-minute, ..., to 20-minute, and one more data set with 3-minute fitting window. The comparison (not shown, similar to Figure \ref{fig:fitcomp}) suggests 4-minute fitting window could be better, but these data sets generally show less uniform distributions when compared with data sets derived from high time resolution data. The same $\chi_{\nu}^2$ and $\Delta(B_r/B)^2$ values are applied to remove bad fittings.

In conclusion, for E1 data from Oct. 20\textsuperscript{th}, 2018 to Nov. 24\textsuperscript{th}, 2018, (1) the temperature anisotropy data are relatively better fitted with 1-minute fitting window and 10-second moving step for Encounter mode data, and 4-minute fitting window and 1-minute moving step for Cruise mode data; (2) the $\chi_{\nu}^2$ smaller than 1 and $\Delta(B_r/B)^2$ larger than 0.05 are necessary to polish the fittings.

\section{Temperature uncertainty estimations} \label{sec:error}
\citet{kasper-2006} presents the first uncertainty estimations for parallel and perpendicular temperature with a technique that is independent of the method used to extract an estimate of the anisotropy from the raw data. They found that the maximum parallel temperature uncertainty occurs predominantly when the magnetic field is out of the ecliptic plane or perpendicular to the Sun-Earth line, and the perpendicular temperature is poorly constrained when the magnetic field is radial. We note that their method relies on there being a statistically significant sample of observations at high plasma beta above 10, which is not the case for the PSP dataset. However, it is valuable to estimate the uncertainty based on our linear fitting method, and check if our method could capture the systematic distribution features.

\begin{figure}
\epsscale{0.75}
\plotone{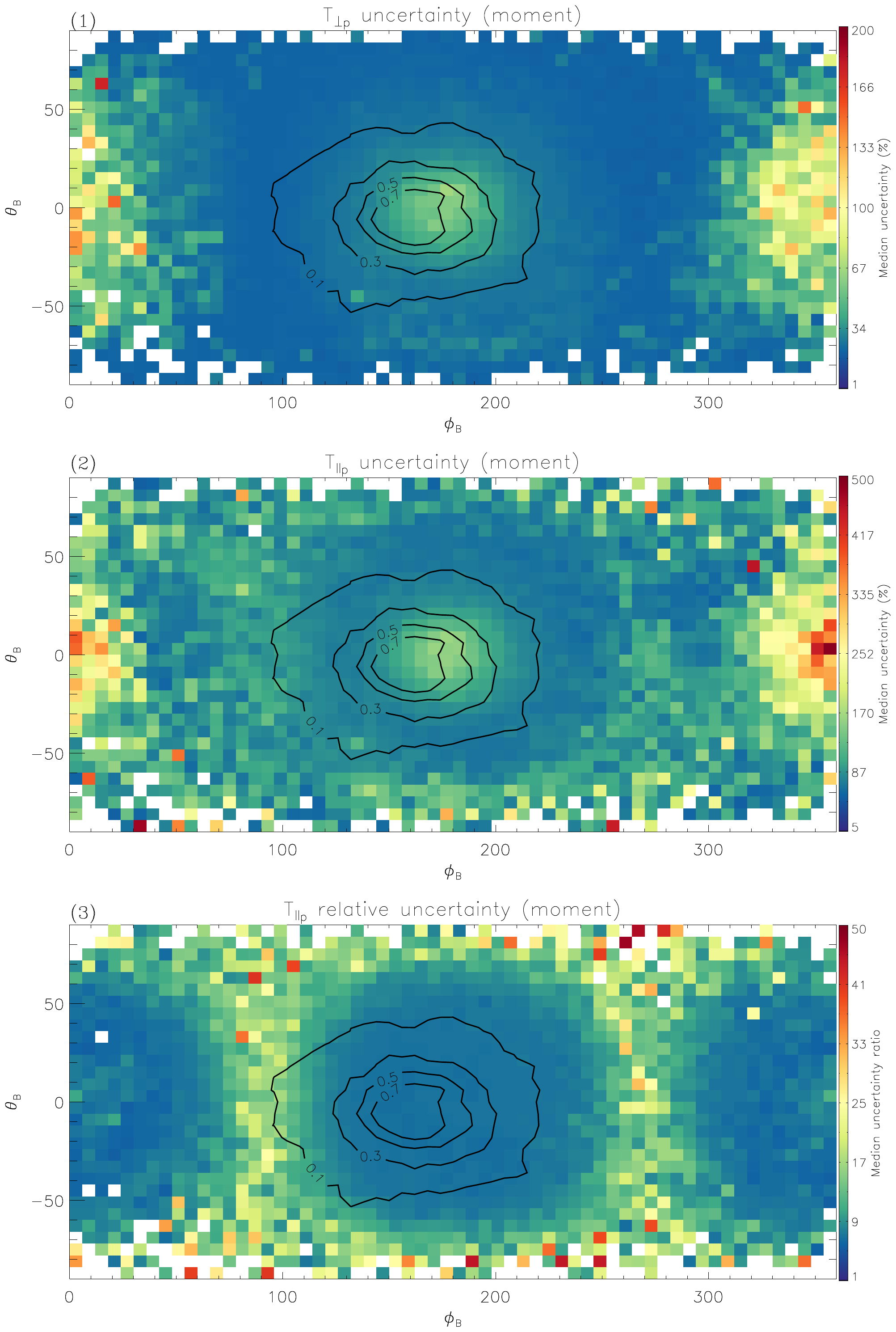}
\caption{The uncertainty distributions of temperature components. Panel (1) and (2) show perpendicular and parallel temperature uncertainty distributions with azimuthal angle $\phi_B$ and elevation angle $\theta_B$, respectively. Panel (3) highlights parallel temperature uncertainty by comparing parallel temperature uncertainty in Panel (2) with perpendicular temperature uncertainty in Panel (1). The black lines indicate 10\%, 30\%, 50\% and 70\% measurement contours.}. \label{fig:tanierror}
\end{figure}

According to Equation \ref{eq:wtilde}, the linear fittings will return 1-sigma uncertainty for both ($T_{\parallel p}-T_{\perp p}$) and $T_{\perp p}$. Therefore, we can directly estimate the uncertainty for $T_{\perp p}$, but need to calculate the propagated uncertainty for $T_{\parallel p}$, implying its uncertainty would partially include contributions from $T_{\perp p}$. Figure \ref{fig:tanierror} shows the temperature uncertainty distributions with azimuthal angle $\phi_B$ and elevation angle $\theta_B$. The black lines indicate 10\%, 30\%, 50\% and 70\% measurement contours, and it clearly shows that most of the measurements are clustered at around $\phi_B \sim 160^o$, which is nearly the Parker spiral angle of magnetic field at PSP's location. Panel (1) shows the $T_{\perp p}$ median uncertainty distributions, with the mean value to be 15.7\% and median value of 7.4\%, and about 15\% of the data have an uncertainty larger than 50\%. It is consistent with the \citet{kasper-2006} result that large $T_{\perp p}$ uncertainty mainly occurs in radial magnetic field direction. Moreover, the uncertainty is generally larger in outward direction than in inward direction, which may be caused by spikes in outward direction \citep{Kasper-2019}. Panel (2) presents the $T_{\parallel p}$ median propagated uncertainty distributions, with the mean value to be 82.0\% and median value of 47.4\%, and about 48\% of the data have an uncertainty larger than 50\%. The overall larger $T_{\parallel p}$ uncertainty is partly contributed by $T_{\perp p}$ uncertainty, and it is difficult to accurately calculate the propagated uncertainty of temperature anisotropy. In order to highlight the $T_{\parallel p}$ uncertainty distributions, we compare the ratio between $T_{\parallel p}$ uncertainty and $T_{\perp p}$ uncertainty in Panel (3), which indicates the same distributions as suggested by \citet{kasper-2006}. The consistency of temperature uncertainty distributions between the two methods, which is not a direct comparison between the results derived from both methods as we cannot apply \textit{non-linear} method to SPC data, imply that our method is reliable within the window of provided uncertainties. Besides, we can further exclude fittings with large uncertainties according to the uncertainty distributions, even though only a small fraction of the measurements have large uncertainties as the contours suggest.

\section{Temperature anisotropy comparisons} \label{sec:compare}
\begin{figure}
\epsscale{1.0}
\plotone{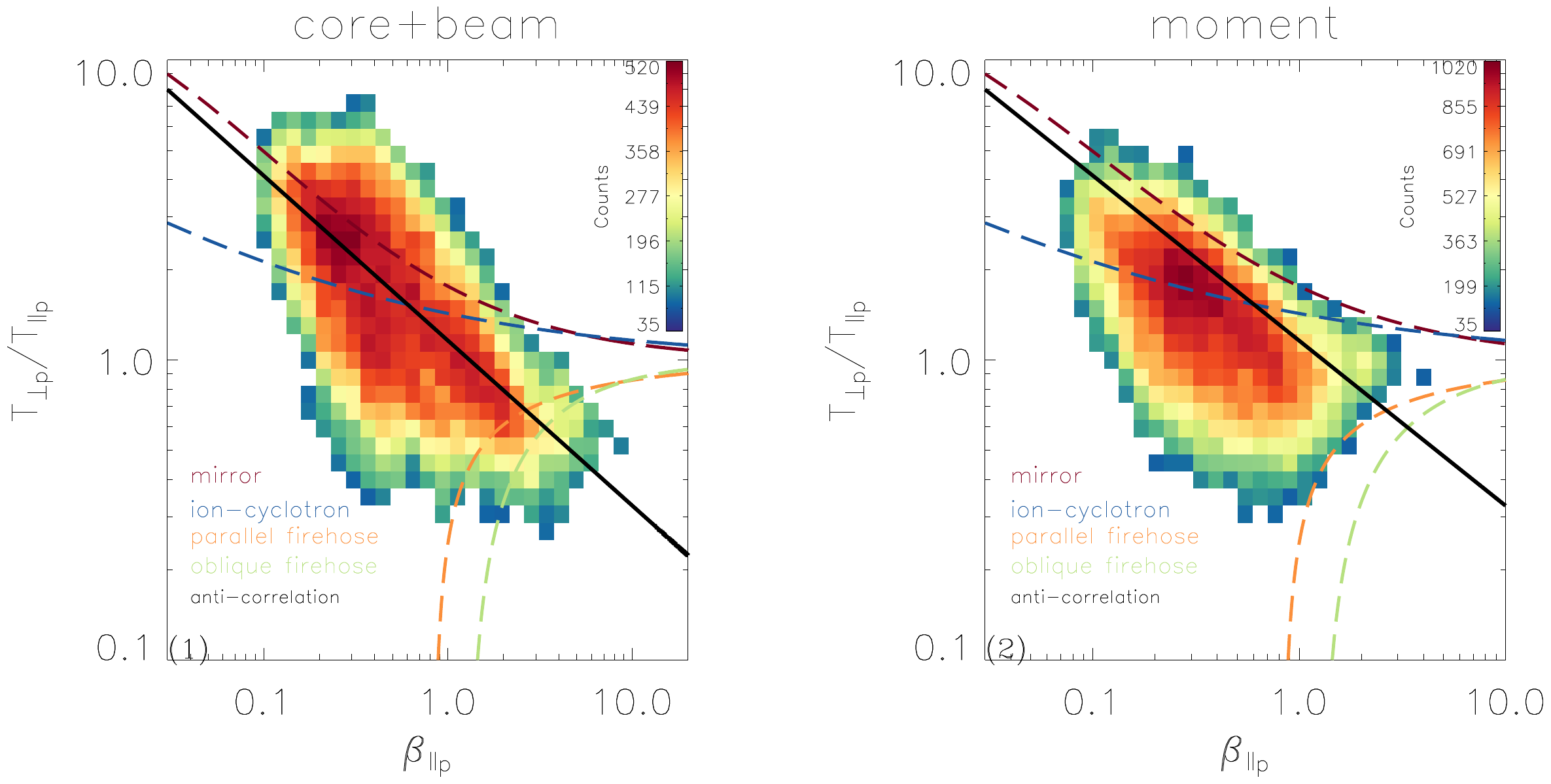}
\caption{Temperature anisotropy distributions. Panel (1) and (2) show plots for core+beam and moment population, respectively. }. \label{fig:alltani}
\end{figure}

For proton data, core, beam, core+beam and moment data are extracted from SPC measurements \citep{Case-2019}, and we fitted the temperature anisotropy for all of them with the method described above. Figure \ref{fig:alltani}, however, only shows the temperature anisotropy versus parallel plasma beta plots for core+beam and moment in Panel (1) and Panel (2), respectively. We note that the instability constraints are calculated for the total proton population assuming the temperature anisotropy can be described as a single bi-Maxwellian velocity distribution function, while it is hard to derive a single set of instability for combined core and beam populations due to the large numbers of multi-ion parameters \citep{matteini-2013}. Thus, we just repeat the same single bi-Maxwellian instability constraints in the core+beam plots for reference under the assumption that we can calculate a total effective parallel and perpendicular temperature by combining the individual fits to the core and beam. Panel (2) shows the overall distributions, which are a combination of slow and fast solar wind as shown in Figure \ref{fig:tanibeta}. The distributions for core+beam population in Panel (1) are expected to be similar to the moment results, which do not separate proton populations. We also note that the core+beam temperature components are derived with our linear fitting method from core+beam dataset, which provides the effective total temperature by merging best fits of core and beam temperatures plus a term due to their relative drift\footnote{ The core+beam effective thermal speed is calculated with equation:
\begin{eqnarray*}
 w_{core+beam}^2 & = & w_{core}^2 \frac{n_{core}}{n_{core}+n_{beam}} +  w_{beam}^2 \frac{n_{beam}}{n_{core}+n_{beam}} + (\overrightarrow{\mathbf{v}}_{core}-\overrightarrow{\mathbf{v}}_{beam})^2 \frac{n_{core} n_{beam}}{(n_{core}+n_{beam})^2}
\end{eqnarray*}}.
However, it seems the moment temperature anisotropies are more physical than that of the core+beam population, which may include systematic errors on parallel temperature estimations due to the core fits not working at low $B_r$. The core population shows similar distributions as the core+beam population, partly because beam seems to be only several percent (median value is 6.16\%) of the core density and they do not yet appear to be reliable. This is one reason why we primarily show the moment results in this work. \citet{kasper-2006} suggests that the temperature components derived with moment method could deviate from \textit{non-linear} results by about 23\%. But we cannot apply \textit{non-linear} fitting method to SPC data due to SPC measures one dimensional reduced VDF, and moment method could be a choice to deal with a single ion spectrum. Besides, we think if anything the moment uncertainties on PSP/SPC should be lower because the VDF is wider generally and resolved into more energy windows. The more gradually the VDF varies over the energy windows of the plasma instrument and the more windows the VDF is detected in at high signal to noise, the more accurate the resulting moments. In addition, we think the moment-derived temperature values could be more physical because they do not impose the assumption of a Maxwellian shape on the velocity distribution functions. Other studies of the proton velocity distribution functions seen by PSP are reporting a significant non-Maxwellian kurtosis near E1 perihelion \citep{Martinovic-2019} , and other significant deviations from the two-Maxwellian model \citep{Case-2019}. This implies that different model functions other than bi-Maxwellian (for example, \citet{wilson-2019a, wilson-2019b} found bi-Kappa works better to fit electron halo and strahl VDFs) might be considered in the future work.

\begin{figure}
\epsscale{1.0}
\plotone{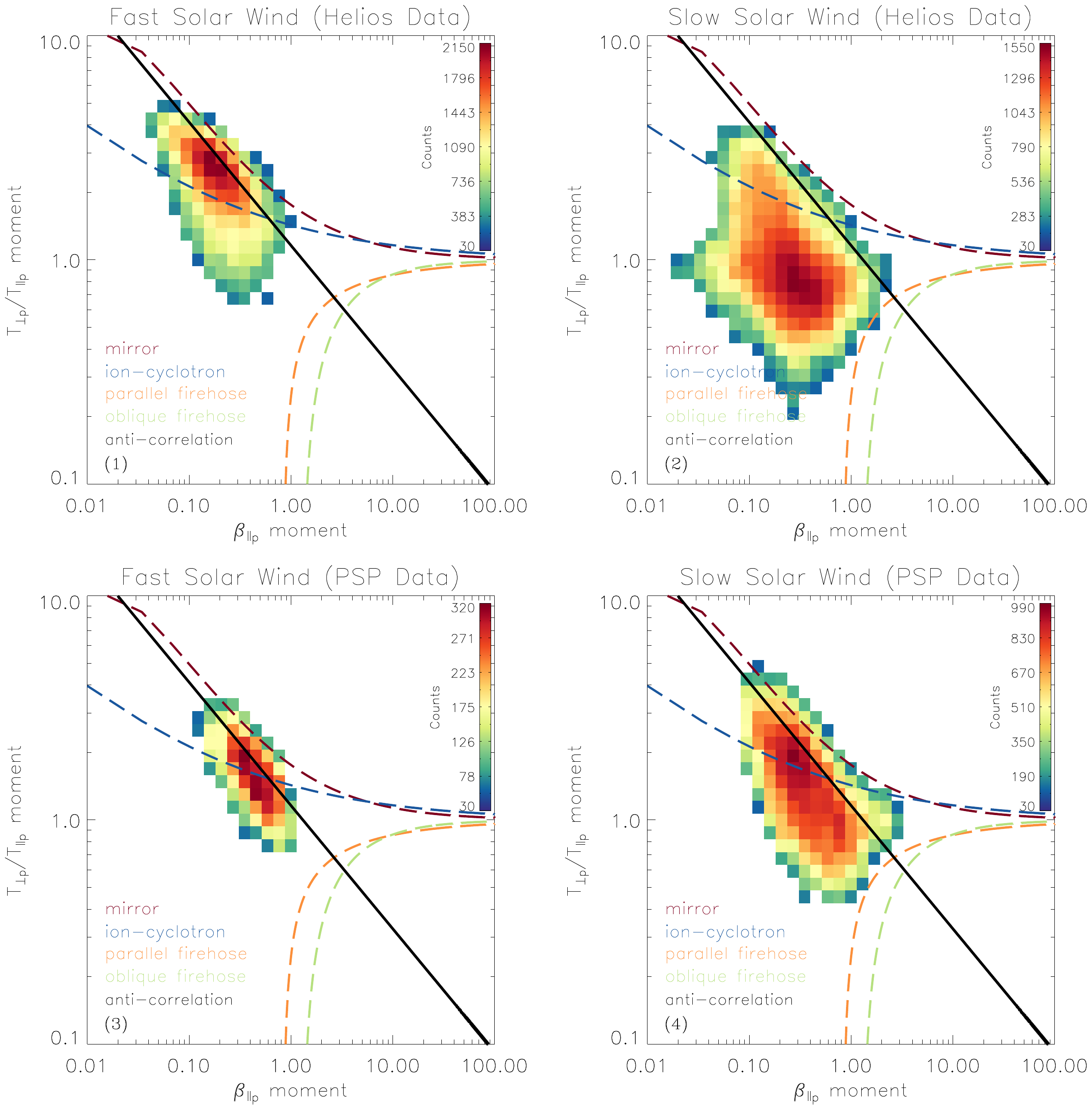}
\caption{Comparison of temperature anisotropy distributions in different solar winds. Panel (1) and Panel (2) show fast solar wind and slow solar wind distributions with Helios moment measurements from about 0.30 to 0.35 AU. Panel (3) and Panel (4) show fast solar wind and slow solar wind distributions with PSP moment observations inside 0.25 AU during E1. }. \label{fig:helios}
\end{figure}

In Figure \ref{fig:helios}, we compare the temperature anisotropy distributions in fast and slow solar wind between Helios and PSP results, with the instability constraints overlaid for reference. In order to compare with the total proton results, we calculate Helios total proton (moment) parameters with current released core and beam results fitted with \textit{non-linear} method by taking into account drift for parallel component. The Helios data from both Helios spacecraft below 0.35 AU are used to facilitate our comparison with PSP moment results inside 0.25 AU. The Helios 1 data covers late 1974 to 1985 and the Helios 2 data covers 1976 to 1980. Panel (1) and (3) indicate strong anisotropies in fast wind measured by both spacecraft, and they match well with the anti-correlation line. Further, it seems Helios observe higher temperature anisotropies in smaller parallel beta region than PSP does. The lower PSP temperature anisotropies could be caused by the PSP observes fast wind with lower speed than Helios, and larger parallel beta may associated with higher parallel temperature than expected in inner heliosphere. An additional explanation for the higher values of parallel beta seen by PSP than Helios is that the overall strength of the coronal magnetic field has been decreasing over the last fifty years \citep{richardson-2002, janardhan-2015}. However, both spacecraft observe the temperature anisotropies show a trend to mirror thresholds. Moreover, the PSP slow wind (Panel (4)) is more anisotropic than the Helios ones (Panel (2)). The difference in slow wind observations may be a consequence of stronger perpendicular temperature heating inside 0.24 AU, or the limited data of the PSP observations.

\begin{figure}
\epsscale{0.75}
\plotone{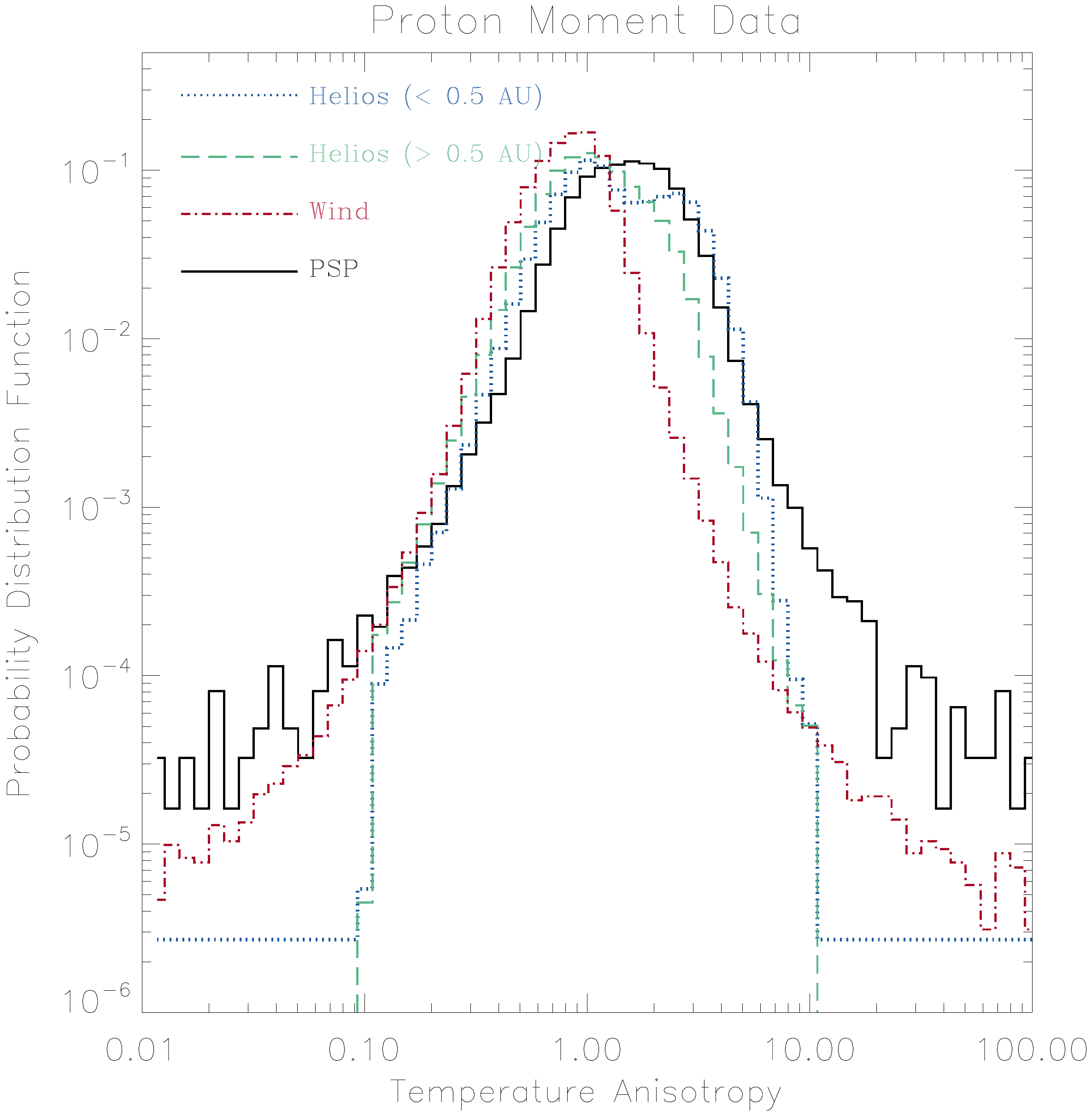}
\caption{Probability distributions of proton moment temperature anisotropy with different spacecraft observations. The black solid line, red dash-dot line, blue dotted line and green dashed line represent PSP data, Wind data, Helios data inside 0.5 AU and Helios data outside 0.5 AU, respectively. }. \label{fig:Tanihist}
\end{figure}

In Figure \ref{fig:Tanihist}, we further compare the probability distribution functions (PDFs) of temperature anisotropies measured by PSP, Helios and Wind spacecraft. Here, we use the Helios total proton data from about 0.3 AU to 1 AU with data described above. The Wind data are collected at 1 AU since June 2004 with Wind/SWE Faraday cups \citep{ogilvie-1995}, and the temperature anisotropies are also derived with\textit{ non-linear }method. The red dash-dot line indicates the Wind anisotropies, with the median value to be 0.78 and mean value is 0.83. In comparison, PSP moment shows near two times larger temperature anisotropies when close to the Sun, with the median value and mean value to be 1.39 and 1.61, respectively. This could be reasonable if the solar wind experiences stronger anisotropic heating closer to the Sun \citep{chew-1956, chandran-2011, hellinger-2011, kasper-2017, Stansby-2019}. Moreover, we compare them with the Helios results inside (blue dotted line) and outside (green dashed line) 0.5 AU. The distributions indicate one peak characterized by isotropic signature that independent with distances, which could be contributed predominantly by slow solar wind. In contrast to the Wind results, Helios observations are bimodal and the second peak is more significant at small radial distances, which is consistent with the temperature anisotropy evolution of fast solar wind \citep[e.g.][]{matteini-2013}. Thus, the PDF distribution of PSP temperature anisotropies seem to be an average of both peaks, which could be a consequence of two reasons. On one hand, the PSP observes rare fast solar wind with speed larger than 600 $km\ s^{-1}$, thus the second anisotropic peak is not that significant. On the other hand, PSP observes solar wind in inner heliosphere may experience stronger perpendicular heating, so the isotropic peak moves to larger values. Besides, the similar PDF distributions between PSP and Wind/Helios results may further suggest that the PSP moment results are reasonable.

\acknowledgments

The SWEAP Investigation and this publication are supported by the PSP mission under NASA contract NNN06AA01C. The SWEAP team expresses its gratitude to the scientists, engineers, and administrators who have made this project a success, both within the SWEAP institutions and from NASA and the project team at JHU/APL. The FIELDS experiment was developed and is operated under NASA contract NNN06AA01C. The Helios proton core and beam data are available on the Helios data archive \url{http://helios-data.ssl.berkeley.edu/}, and the Wind data come from CDAWeb/SPDF \url{ftp://spdf.gsfc.nasa.gov/pub/data/wind/}. D.V was supported by NASA's Future Investigators in NASA Earth and Space Science and Technology Program Grant 80NSSC19K1430. M.M.M is supported by NASA grant 80NSSC19K1390. K.G.K is supported by NASA grant 80NSSC19K0912.




\bibliography{TAnisotropy}{}
\bibliographystyle{aasjournal}


\end{document}